\documentclass[twocolumn,showpacs,amsmath,amssymb,superscriptaddress]{revtex4-1}

\usepackage{graphicx}
\usepackage{dcolumn}
\usepackage{bm}
\usepackage[abs]{overpic}
\usepackage{here}
\usepackage{color}
\usepackage{url}
\usepackage{accents}
\def\up{\uparrow}
\def\down{\downarrow }
\def\Vec#1{\bm{#1}}

\begin{document}


\title{Time-reversal symmetry breaking and gapped surface states due to spontaneous emergence of new order in $d$-wave nanoislands}
\author{Yuki Nagai}
\affiliation{CCSE, Japan  Atomic Energy Agency, 178-4-4, Wakashiba, Kashiwa, Chiba, 277-0871, Japan}
\affiliation{Department of physics, Massachusetts Institute of Technology, Cambridge, MA 02139, USA}

\author{Yukihiro Ota}
\affiliation{Research Organization for Information Science and Technology (RIST), 1-5-2 Minatojima-minamimachi, Kobe, 650-0047, Japan}

\author{K. Tanaka}
\affiliation{Department of Physics and Engineering Physics, University of Saskatchewan, 116 Science Place, Saskatoon, SK. S7N 5E2, Canada}

\date{\today}
             
\begin{abstract}
We solve the Bogoliubov-de Gennes equations self-consistently for the $d$-wave order parameter in nanoscale $d$-wave systems with [110] surfaces and show that spontaneous time-reversal symmetry (TRS) breaking occurs at low temperatures due to a spontaneously induced complex order parameter of extended $s$-wave symmetry. The Andreev surface bound states, which are protected by a one-dimensional (1D) topological invariant in the presence of TRS, are gapped by the emergence of this new order parameter.
The extended $s$-wave order parameter is localized within a narrow region near the surfaces, which is consistent with the fact that topological protection of the gapless Andreev surface states is characterized by the 1D topological invariant. 
In this TRS-breaking phase, not only is the complex order parameter induced, but also the $d$-wave order parameter itself becomes complex.
Furthermore, the disappearance of topological protection brings about novel vortex phenomena near the surfaces. We show that vortex-antivortex pairs are formed in the extended $s$-wave order parameter along the surfaces if the side length of a nanoisland or the width of an infinitely long nanoribbon is relatively large.
\end{abstract}

\maketitle

Symmetry is at the heart of both high-energy and condensed matter physics. Symmetry breaking underlies not only mass in our universe, but also superconductivity in metals. Symmetry classifies topological materials \cite{Schnyder}. Symmetry protects topological order, such as the Haldane phase in a spin-1 chain with time-reversal, parity, and translation symmetries \cite{GuSPT}. 

Time-reversal symmetry (TRS) and topological phenomena associated with it are hot topics in condensed matter physics. Sato {\it et al.} have shown the bulk-edge correspondence between the zero-energy Andreev bound states on [110] surfaces of a high-$T_{\rm c}$ cuprate superconductor and a topological invariant protected by TRS \cite{Satoflat}. 
The Andreev bound states are understood as arising from the sign-changing property of the $d_{x^{2}-y^{2}}$-wave order parameter \cite{Tsuei,Kashiwaya}.
From the topological point of view, such zero-energy surface states are robust as the bulk-edge correspondence ensures the existence of gapless surface states in a topologically nontrivial system \cite{Hasan,Alicea}.  
In a 2D nodal $d$-wave superconductor, a 1D topological invariant as a function of momentum along the [110] surface can be defined, through which TRS provides topological protection of the Andreev bound states \cite{Satoflat}.

Gapless surface states can become gapped when the corresponding symmetry is broken. 
The surface states in a topological insulator acquire a gap as a result of spontaneous or non-spontaneous breaking of TRS \cite{Chen,Yu}.
An example of spontaneous TRS breaking is the anomalous Hall effect in tetradymite semiconductors, where the ferromagnetic order parameter induced by the Van Vleck type spin susceptibility results in the gapped surface states \cite{Yu}.
Splitting of the Andreev bound states due to spontaneously induced surface currents, i.e., spontaneous TRS breaking has been discussed by assuming the existence of a subdominant order parameter with relative phase $\pi/2$, which is stabilized on the [110] surface of a $d$-wave superconductor \cite{Kashiwaya,Matsumoto,Fogelstroem,note}. This has been a controversial topic in the past two decades owing to contradictory experimental results on high-$T_{\rm c}$ cuprates \cite{Gustafsson,Hakansson}. A recent experiment, however, has clearly detected a full gap in the excitation spectrum that is consistent with spontaneously broken TRS in a nanoscale YBa$_{2}$Cu$_{3}$O$_{7-\delta}$ island \cite{Gustafsson}.
By introducing an intrinsic $s$-wave pairing interaction and controlling its strength in terms of the ratio of the bare bulk transition temperature to that of the $d$-wave order, Black-Schaffer {\it et al.}\cite{Black} have found that the resulting $d+is$ symmetry is consistent with the experimental findings of Ref.~\onlinecite{Gustafsson}.

Vorontsov has proposed that films of a $d$-wave superconductor at low temperatures can exhibit unusual superconducting phases due to transverse confinement, with TRS or continuous translational symmetry broken spontaneously \cite{Vorontsov}.
Recently, a theoretical study of $d$-wave superconducting nanoislands within the quasiclassical Eilenberger framework with no subdominant pairing channel has predicted spontaneous breaking of TRS at low temperatures, where staggered fractional vortices are formed along the surfaces, each containing the zero-energy Andreev bound states at its center \cite{Hakansson}. The quasiclassical formulation, however, is limited in that it cannot resolve individual quasiparticle excitations and it is not a good approximation for high-$T_{\rm c}$ cuprates as it requires the coherence length to be much larger than $1/k_F$.

\begin{figure}[t]
\begin{center}
     \begin{tabular}{p{ 0.8 \columnwidth}} 
      \resizebox{0.8 \columnwidth}{!}{\includegraphics{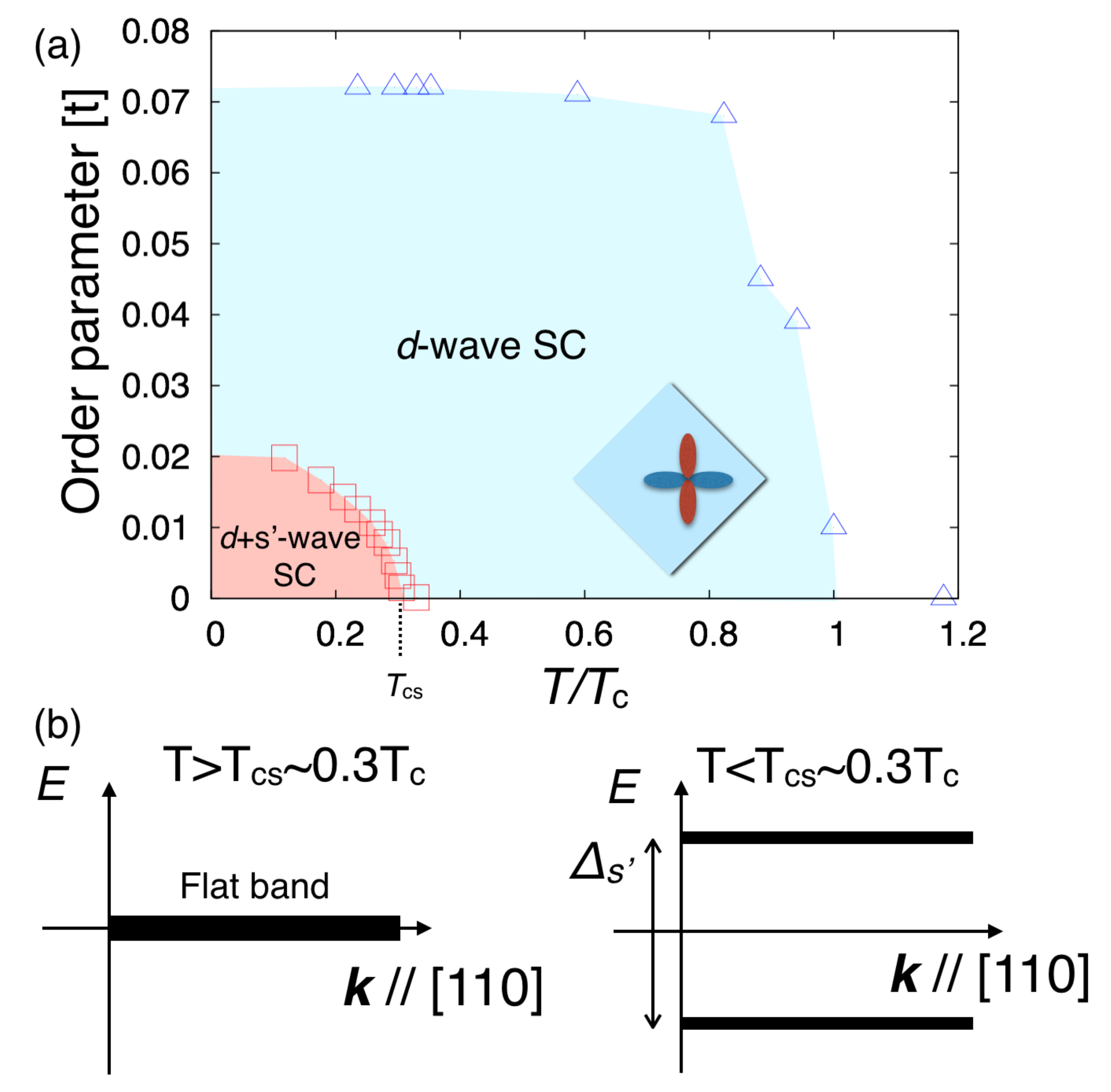}} 
    \end{tabular}
\end{center}
\caption{
(Color online) (a) Temperature dependence of the order parameters in the $d$-wave nanoisland. 
The system size is $L_{x} \times L_{y} = 96 \times 96$ and the critical temperature is $T_{\rm c} = 0.085t$. (b) Schematic illustration of the eigenvalues around zero energy. 
The zero-energy Andreev bound states on the [110] surface are gapped because of spontaneous time-reversal symmetry breaking. The gap size is characterized by the extended $s$-wave order parameter that is induced along the surfaces.
\label{fig:fig1}
 }
\end{figure}

In this work, by solving the Bogoliubov-de Gennes (BdG) equations \cite{deGennes} self-consistently for the $d$-wave order parameter in $d_{x^{2}-y^{2}}$-wave nanoislands with [110] surfaces, we show that spontaneous TRS breaking occurs at low temperatures due to {\it spontaneous} emergence of a new complex order parameter. The spontaneous disappearance of topological protection is accompanied by this new order parameter that is induced only near the surfaces below a certain temperature, which is lower than the superconducting transition temperature $T_{\rm c}$. This order parameter has extended $s$-wave symmetry and it characterizes the energy gap of the split Andreev bound states on the surfaces, as schematically illustrated in Fig.~\ref{fig:fig1}(b) \cite{Sup}. This is analogous to the gap opening on surfaces of a topological insulator originating from the spontaneously induced ferromagnetic order \cite{Yu}. We find that the phase transition is of second order, similarly to the transition to the TRS-breaking state within the purely $d$-wave phase found by the quasiclassical Eilenberger approach \cite{Hakansson}. Our approach beyond the quasiclassical theory allows us to accurately study the discrete spectrum of quasiparticle excitation, unraveling a close relation between the spontaneously induced order parameter and the surface excitation. The extended $s$-wave order parameter is localized within a narrow region along the surfaces: this is compatible with the fact that the topological protection is characterized by a 1D topological invariant defined along the surfaces.
We also find that when the phase transition to this TRS-breaking phase occurs, the $d$-wave order parameter itself becomes complex.

To summarize, spontaneous disappearance of topological protection occurs in a $d_{x^{2}-y^{2}}$-wave nanoisland with [110] surfaces as a second-order phase transition, leading to the gapped surface excitation originating from the formation of extended $s$-wave Cooper pairs near the surfaces. 
This is an example of a spontaneously induced order parameter being behind the breaking of the symmetry that would otherwise protect the gapless surface states.
Furthermore, the disappearance of topological protection brings about novel vortex phenomena near the surfaces. We show that vortex-antivortex pairs are formed along the surfaces of a $d_{x^{2}-y^{2}}$-wave nanoisland when the side length is relatively long. The phase winding occurs in the extended $s$-wave order parameter, where there is a $2\pi$ phase winding around each vortex or antivortex. 
It is shown in Supplemental Material \cite{Sup} that the disappearance of zero-energy surface states due to spontaneous TRS breaking can be understood in terms of the fundamental group \cite{Nakahara} for homogeneous systems with spin-singlet pairing. The fundamental group corresponding to the ground-state manifold changes from $\pi_{1}(S^{1})=\mathbb{Z}$ to $\pi_{1}(S^{2})=0$ as the TRS-breaking order emerges \cite{SatoB,Sup}.

We consider the tight-binding model Hamiltonian ${\cal H} = {\cal H}_{\rm BCS} + {\cal H}_{\rm fab}$. 
Here the generalized Bardeen-Cooper-Schrieffer (BCS) Hamiltonian for $d$-wave superconductivity is ${\cal H}_{\rm BCS} = \sum_{ij,\sigma} (-t_{ij} - \mu) c_{i \sigma}^{\dagger} c_{j \sigma} + \sum_{ij} \left[ 
\Delta_{ij} c_{i \up}^{\dagger} c_{j \down}^{\dagger} + {\rm H.c.} 
\right]$, where $c_{i \sigma}^{\dagger}$ creates the electron with spin $\sigma$ at site $i$ and $\mu$ denotes the chemical potential. 
The hopping is restricted within nearest-neighbor sites for simplicity. 
We use the unit system with $\hbar = k_{\rm B} = 1$. 
The Hamiltonian for the fabrication potential ${\cal H}_{\rm fab}$ is written as ${\cal H}_{\rm fab} = \sum_{i,\sigma} V_{i} c_{i \sigma}^{\dagger} c_{i \sigma}$, where the potential $V_{i} = 0$ (500$t$) inside (outside) the nanoisland. 
One can diagonalize ${\cal H}$ to solve the BdG equations expressed as 
\begin{align}
\sum_{j} \left(\begin{array}{cc}
\left[ \hat{H}^{\rm N} \right]_{ij} & [\hat{\Delta}]_{ij} \\
\left[ \hat{\Delta}^{\dagger} \right]_{ij}  & -\left[ \hat{H}^{\rm N \ast} \right]_{ij}
\end{array}\right)
\left(\begin{array}{c}
u_{\gamma}(\Vec{r}_{j}) \\
v_{\gamma}(\Vec{r}_{j}) 
\end{array}\right)
&= 
E_{\gamma}
\left(\begin{array}{c}
u_{\gamma}(\Vec{r}_{i}) \\
v_{\gamma}(\Vec{r}_{i}) 
\end{array}\right). \label{eq:bdg}
\end{align}
Here $\left[ \hat{H}^{\rm N} \right]_{ij} = - t_{ij} - (\mu - V_{i}) \delta_{ij}$ and $[\hat{\Delta}]_{ij} = V_{ij} \sum_{\gamma=1}^{2N} u_{\gamma}(\Vec{r}_{j}) v_{\gamma}^{\ast}(\Vec{r}_{i}) f(E_{\gamma})$, 
where $N$ is the number of lattice sites, $V_{ij}$ denotes the pairing interaction, and $f(x)$ is the Fermi-Dirac distribution function. 
For the results presented below, the chemical potential is fixed to be $\mu = -1.5t$ and the pairing interaction is nonzero only between nearest-neighbor sites, 
$V_{ij} \equiv U = -2t$, so that the resulting order parameter is purely of the $d_{x^{2}-y^{2}}$-wave symmetry.
The order parameter at site $i$ is given in terms of the $d$-wave mean fields by
\begin{equation}
\Delta_{d,i} =  (\Delta_{\hat{x},i}+\Delta_{-\hat{x},i} - \Delta_{\hat{y},i} - \Delta_{-\hat{y},i})/4 \label{eq:dwave}
\end{equation}
with $\Delta_{\pm \hat{e},i} = \Delta(\Vec{r}_{i},\Vec{r}_{i} \pm \hat{e})$, where $\hat{x}$ and $\hat{y}$ denote the unit vectors in the square lattice. 
Considering two-dimensional systems, we assume that the field penetration depth $\lambda$ is infinity.
We use pure $d$-wave mean fields with randomly-distributed phases as the initial guess for our self-consistent calculation.

\begin{figure*}[t]
\begin{center}
     \begin{tabular}{p{ 2 \columnwidth}} 
      \resizebox{2 \columnwidth}{!}{\includegraphics{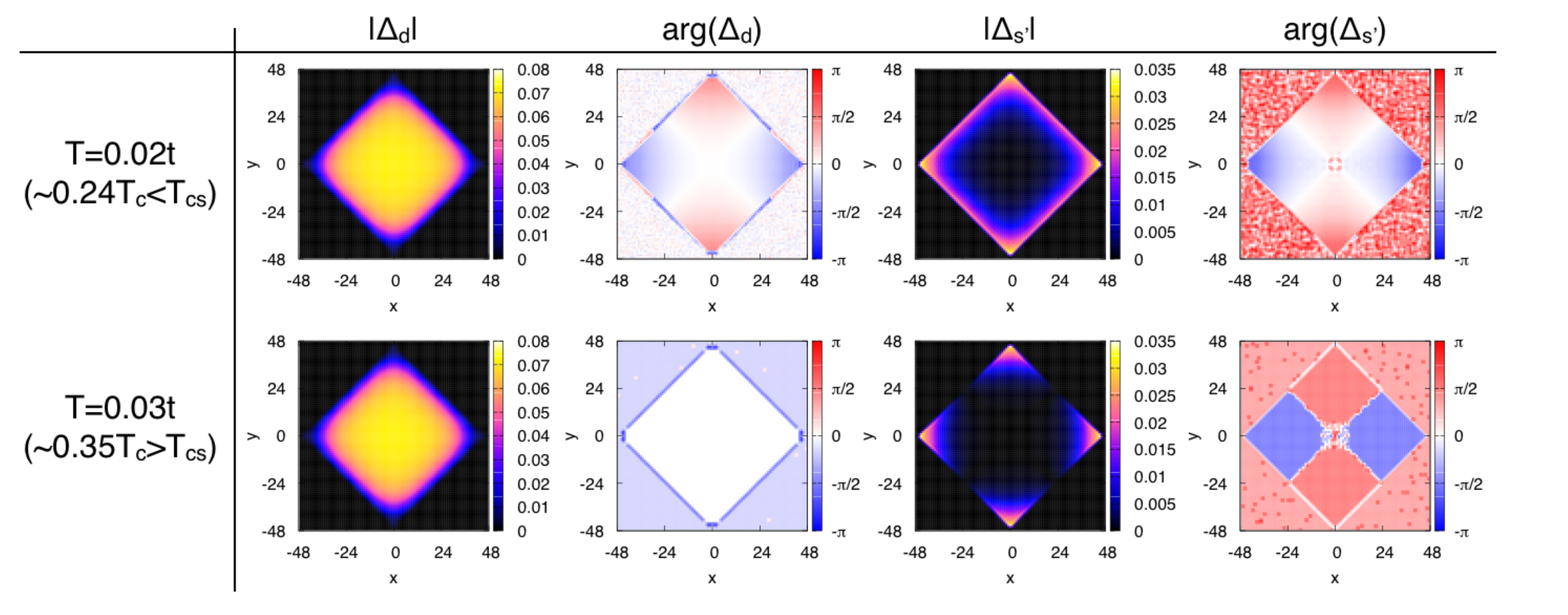}} 
    \end{tabular}
\end{center}
\caption{
(Color online) $d$-wave and extended $s$-wave order parameters below and above $T_{\rm cs}$. The system size is $L_{x} \times L_{y} = 96 \times 96$. To clearly present the spatial variation, arg($\Delta_{s^\prime}$) has been shifted by $\pi/2$.
\label{fig:fig2}
 }
\end{figure*}

The BdG equations (\ref{eq:bdg}) are solved self-consistently along with the $d$-wave gap equation with the use of the reduced-shifted-conjugate-gradient (RSCG) method \cite{NagaiRSCG}.  
The mean field $\langle c_{i} c_{j} \rangle$ can be expressed as $\langle c_{i} c_{j} \rangle = T \sum_{n=-n_{\rm c}}^{n_{c}} \Vec{e}(j)^{\rm T} \Vec{x}(i,\omega_{n})$, 
where $[\Vec{e}(i)]_{\gamma} = \delta_{i \gamma}$ and $n_{\rm c}$ is the cutoff parameter for the Matsubara frequencies, $\omega_{n} = (2 n +1) \pi T$.
The $2N$-dimensional vector $\Vec{x}(i,\omega_{n})$ is obtained by solving the linear equations defined by
$(i \omega_{n} \hat{I} -\hat{H}_{\rm BdG}) \Vec{x}(i,\omega_{n}) = \Vec{h}(i)$,
with $[\Vec{h}(i)]_{\gamma} = \delta_{i+N \gamma}$ and the BdG matrix $\hat{H}_{\rm BdG}$ on the left-hand side (LHS) of Eq.~(\ref{eq:bdg}).
These linear equations with different frequencies can be solved simultaneously by the RSCG method \cite{NagaiRSCG}. 
We use the maximum Matsubara frequency $\omega_{\rm c} = 240 \pi = \pi T (2 n_{\rm c}+1)$. 
A criterion for the residual is set to 0.1, which is small enough to obtain accurate Green's functions \cite{NagaiRSCG}.

To find the true ground state on the mean-field level, 
once self-consistency has been achieved, we calculate the thermodynamic potential given as \cite{Kosztin,Hosseini} 
\begin{align}
\Omega_{s} &= - T \sum_{\gamma=1}^{2N} \ln \left[ 1 + \exp \left(\frac{E_{\gamma}}{T} \right) \right] - \sum_{ij} \frac{|[\hat{\Delta}]_{ij}|^{2}}{U}.
\end{align}
We obtain all the eigenvalues $E_{\gamma}$ by means of 
the Sakurai-Sugiura (SS) method \cite{Sakurai} after the self-consistent calculation \cite{Sup}. 
The SS method allows us to extract the eigenpairs whose eigenvalues are located in a given domain on the complex plane from a generic matrix \cite{zPares,Futamura,NagaiSS}. 
We find that at a certain temperature, the system goes through a second-order phase transition into a state where the extended $s$-wave order parameter defined by \cite{Franz}
\begin{equation}
\Delta_{s',i} =  (\Delta_{\hat{x},i}+\Delta_{-\hat{x},i} + \Delta_{\hat{y},i} + \Delta_{-\hat{y},i})/4 \label{eq:swave}
\end{equation}
becomes nonzero along the surfaces. This order parameter is complex and results in the state having gapped Andreev bound states and lower energy than that of the pure $d$-wave phase with gapless surface excitation. A complex order parameter produces currents and thus breaks TRS (see discussion in relation to topology in Sec.~S6 \cite{Sup}).
 
Figure \ref{fig:fig1}(a) shows the temperature dependence of the two order parameters in a $d_{x^{2}-y^{2}}$-wave nanoisland with $L_{x} \times L_{y} = 96 \times 96$ lattice sites. 
The $d$-wave order parameter is defined by Eq.~(\ref{eq:dwave}) at the center of the nanoisland. 
The extended $s$-wave order parameter, which we call $s'$-wave,  is defined by the minimum eigenvalue of the converged BdG equations. 
The $d$-wave order parameter suddenly goes to zero at the critical temperature $T_{\rm c} =0.085t$, as the coherence length reaches the system size at this temperature.
The second-order phase transition to the phase with nonzero $s'$-wave order parameter along the surfaces occurs at $T/T_{\rm c} \sim 0.3$.
We call this transition temperature $T_{\rm cs}$. 
As can be seen in Fig.~\ref{fig:fig2}, TRS is broken below $T_{\rm cs}$.
The $s'$-wave order parameter is finite around corners of the nanoisland above $T_{\rm cs}$, as it can be induced by a scatterer such as a nonmagnetic impurity in a $d$-wave superconductor \cite{Franz}.
This localized $s'$-wave order parameter becomes global at temperature $T_{\rm cs}$, i.e., nonzero all along the surfaces \cite{Sup}. 
Below $T_{\rm cs}$ the $d$-wave order parameter also becomes complex, while the BdG equations converge to a real $d$-wave order parameter above $T_{\rm cs}$ despite the complex initial guess. 
In terms of topology, a 1D order parameter should be induced at the TRS-breaking temperature $T_{\rm cs}$, 
since the topology of nodal time-reversal-invariant superconductors is characterized by a 1D topological invariant \cite{Sup}.
The $s'$-wave order parameter is 1D in the sense that it is induced within a narrow region along the surfaces.
This second-order phase transition can be probed by various kinds of surface-sensitive experiments such as scanning tunneling microscopy or point contact spectroscopy.

\begin{figure}[t]
\begin{center}
     \begin{tabular}{p{ 1 \columnwidth}} 
      \resizebox{1 \columnwidth}{!}{\includegraphics{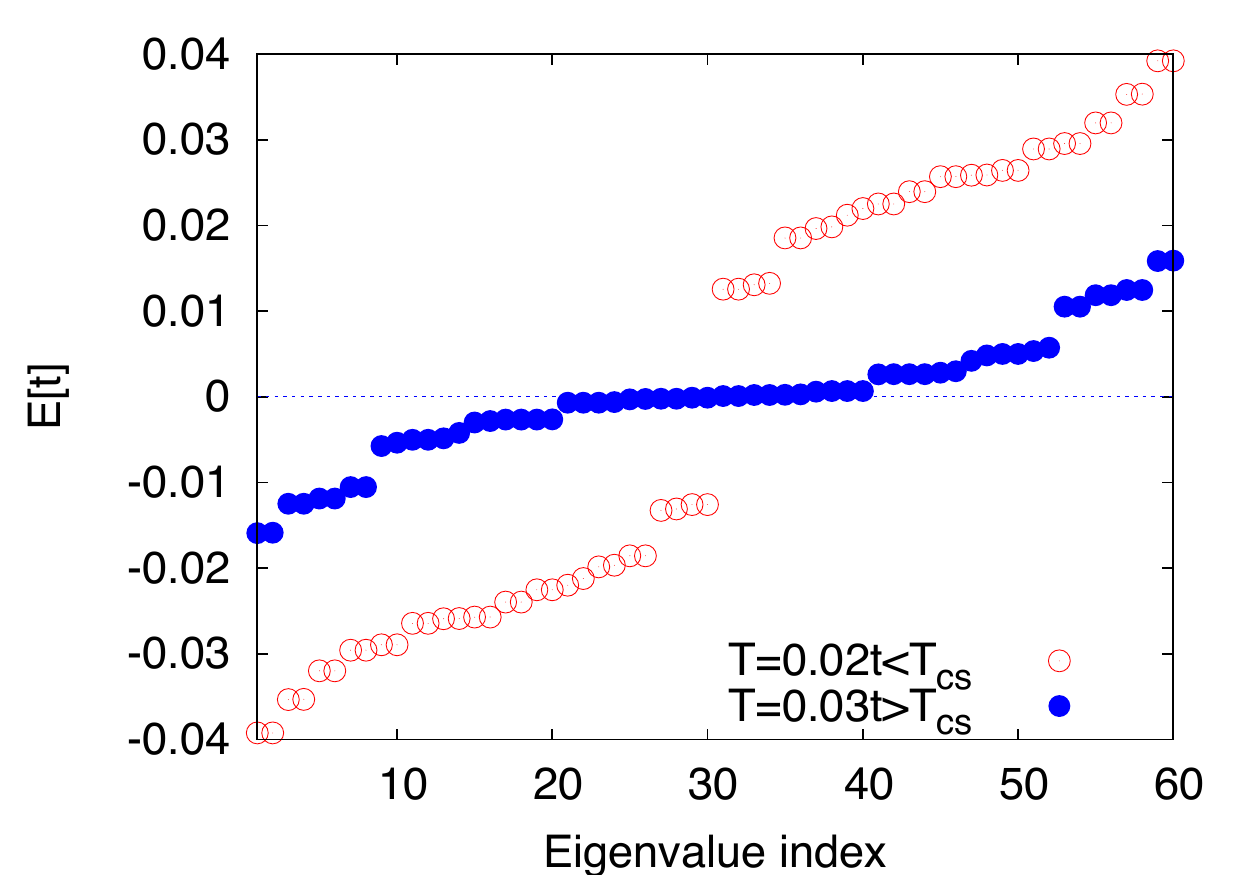}} 
    \end{tabular}
\end{center}
\caption{
(Color online) Eigenvalue distribution below and above the time-reversal-symmetry breaking temperature $T_{\rm cs}$. 
The parameters are the same as for Fig.~\ref{fig:fig1}.
\label{fig:fig3}
 }
\end{figure}

Figure \ref{fig:fig3} shows the eigenvalue distributions below and above $T_{\rm cs}$. 
One can see that the spectrum is gapped below $T_{\rm cs}$. 
In the zero-temperature limit, the thermodynamic potential reduces to 
\begin{align}
\Omega_{s} &\sim -\sum_{\gamma=1}^{2N} E_{\gamma} \theta(E_{\gamma})- \sum_{ij} \frac{|[\hat{\Delta}]_{ij}|^{2}}{U}, \label{eq:energy}
\end{align}
which is equivalent to the internal energy. 
Here $\theta(x)$ is the Heaviside step function. 
Equation (\ref{eq:energy}) clearly shows that the zero-energy eigenvalues of the Andreev bound states do not contribute to the minimization of the internal energy. 
The first term in Eq.~(\ref{eq:energy}) decreases when the flat band in a superconductor is gapped. 
The second term is usually small since the system satisfies the relation $|[\hat{\Delta}]_{ij}|^{2}/|U| < |[\hat{\Delta}]_{ij}|$. 
Thus, the gapped phase can become the ground state. 
In a $d_{x^{2}-y^{2}}$-wave nanoisland, the flat band is spontaneously split as the extended $s$-wave order parameter emerges globally along the surfaces below $T_{\rm cs}$. 
We propose that our findings are rather universal and that
an unconventional superconductor with gapless surface states such as $p_{x}$-wave superconductors \cite{Satoflat} can have a new symmetry-breaking phase with a second order phase transition.

\begin{figure}[t]
\begin{center}
     \begin{tabular}{p{ 1 \columnwidth}} 
      \resizebox{1\columnwidth}{!}{\includegraphics{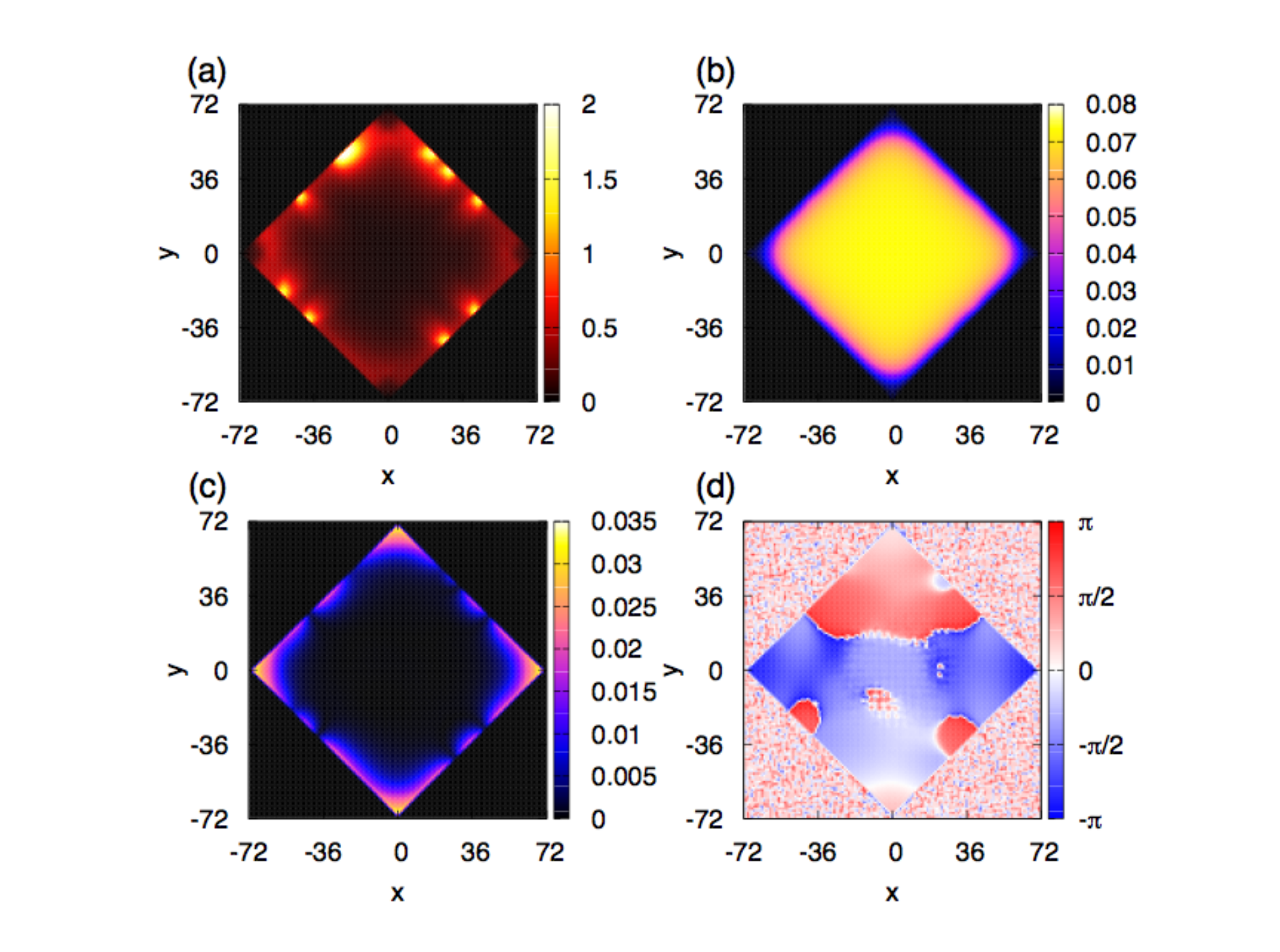}} 
    \end{tabular}
\end{center}
\caption{
(Color online) Results for the larger $d$-wave nanoisland ($L_{x} \times L_{y} = 144 \times 144$) at temperature $T = 0.02t$: (a) zero-energy local density of states, (b) the amplitude of the $d$-wave order parameter, (c) the amplitude of the extended $s$-wave order parameter, (d) the phase of the extended $s$-wave order parameter. 
\label{fig:fig4}
 }
\end{figure}

We find the appearance of vortex-antivortex pairs in relatively large nanoislands. 
In a larger system with $L_{x} \times L_{y} = 144 \times 144$ sites, the TRS-breaking temperature $T_{\rm cs}$ is almost the same as that in the smaller system shown in Fig.~\ref{fig:fig1}(a).
In Fig.~\ref{fig:fig4} we present (a) the zero-energy local density of states, (b) $|\Delta_{d,i}|$, (c) $|\Delta_{s',i}|$, and (d) the phase of $\Delta_{s',i}$ for the $L_{x} \times L_{y} = 144 \times 144$ nanoisland.
There are zero-energy bound states on the surfaces as can be seen in Fig.~\ref{fig:fig4}(a). 
The zero-energy local density of states has been calculated using the eigenfunctions obtained by the SS method, with the Lorentzian smoothing width $\eta = 0.01t$ \cite{Sup}. 
It can be seen in Fig.~\ref{fig:fig4}(b) that
the $d$-wave order parameter is similar to that in the smaller system shown in Fig.~\ref{fig:fig2}. 
Contrary to Fig.~\ref{fig:fig2}, however, the $s'$-wave order parameter in Fig.~\ref{fig:fig4}(c) vanishes at the positions that have zero-energy bound states.
One can see in Fig.~\ref{fig:fig4}(d) that a phase singularity occurs in the $s'$-wave order parameter at each of these positions.
Thus, these zero-energy bound states are the Andreev bound states in a vortex or an antivortex in the induced $s'$-wave order parameter.
Vortices and antivortices can appear in spontaneous TRS-broken 
systems and in the work of Ref.~\cite{Hakansson}, fractional vortex-antivortex chains along the surfaces were found in the $d$-wave order parameter.
In contrast, our TRS-broken phase has {\it integer} vortex-antivortex pairs in the induced extended $s$-wave order parameter, since its phase as shown in Fig.~\ref{fig:fig4}(d) winds from $-\pi$ to $\pi$ or vice versa.
We note that the positions of vortex-antivortex pairs depend on the initial randomly-distributed phase of the $d$-wave order parameter.
We have confirmed that such vortex-antivortex pairs appear in a larger system for $L_{x} \times L_{y} = 192 \times 192$. 

Although we have focused on the chemical potential $\mu=-1.5t$ and the $d$-wave pairing interaction $U=-2t$ in presenting our results, we have found the second-order phase transition to the spontaneous TRS-broken phase with the induced $s'$-wave order for a wide range of $\mu$ and $U$. 
To address the question as to whether spontaneous TRS breaking happens only in nanoislands due to quantum confinement or it is a generic feature of the [110] surface, we have also performed self-consistent calculation for nanoribbons, which have infinitely long [110] surfaces, but a finite width in the perpendicular direction. We have found that the second-order phase transition to the TRS-broken phase where the $s'$-wave order is induced along the surface is generic to [110] surfaces and also occurs in nanoribbons. For $\mu=-1.5t$ and $U=-2t$, vortex-antivortex pairs appear when the width of a nanoribbon or the side length of a nanoisland is roughly eight times or larger than the coherence length. 
Detailed studies of nanoribbons will be presented in a future publication.

In conclusion, we have shown by solving the BdG equations self-consistently for nanoscale $d$-wave systems that spontaneous emergence of new order results in gapping of the zero-energy surface states by breaking the underlying symmetry. 
We have found that not only is the induced order parameter complex, but also the $d$-wave order parameter itself becomes complex in the phase with broken TRS.
The gapped surface states on [110] surfaces in the TRS-broken phase of a $d$-wave superconductor can be detected by surface-sensitive probes, and the Andreev bound states in vortex-antivortex pairs in the induced extended $s$-wave condensate can be distinguished from the gapless surface states in the TRS-preserving phase by measuring the concurrent magnetic field with nanoscale SQUIDs \cite{Hakansson,Vasyukov}.

The authors would like to acknowledge Masahiko Machida, Hiroki Isobe, and Frank Marsiglio for helpful discussions and comments. 
The calculations were performed on the supercomputing system SGI ICE X at the Japan Atomic Energy Agency. 
This study was partially supported by JSPS KAKENHI Grant Number 26800197, the “Topological Materials Science” (No. JP16H00995) KAKENHI on Innovative Areas from JSPS of Japan, and the Natural Sciences and Engineering Research Council of Canada. 


\clearpage
\newpage
\widetext
\onecolumngrid

\onecolumngrid
\setcounter{equation}{0}
\renewcommand{\thefigure}{S\arabic{figure}} 

\setcounter{figure}{0}

\renewcommand{\thesection}{S\arabic{section}.} 
\renewcommand{\theequation}{S\arabic{equation}} 
\renewcommand{\thetable}{S\arabic{table}} 
\begin{flushleft} 
{\Large {\bf Supplemental material}}
\end{flushleft} 

\begin{figure}[ht]
\begin{center}
     \begin{tabular}{p{ 0.5 \columnwidth}} 
      \resizebox{0.5 \columnwidth}{!}{\includegraphics{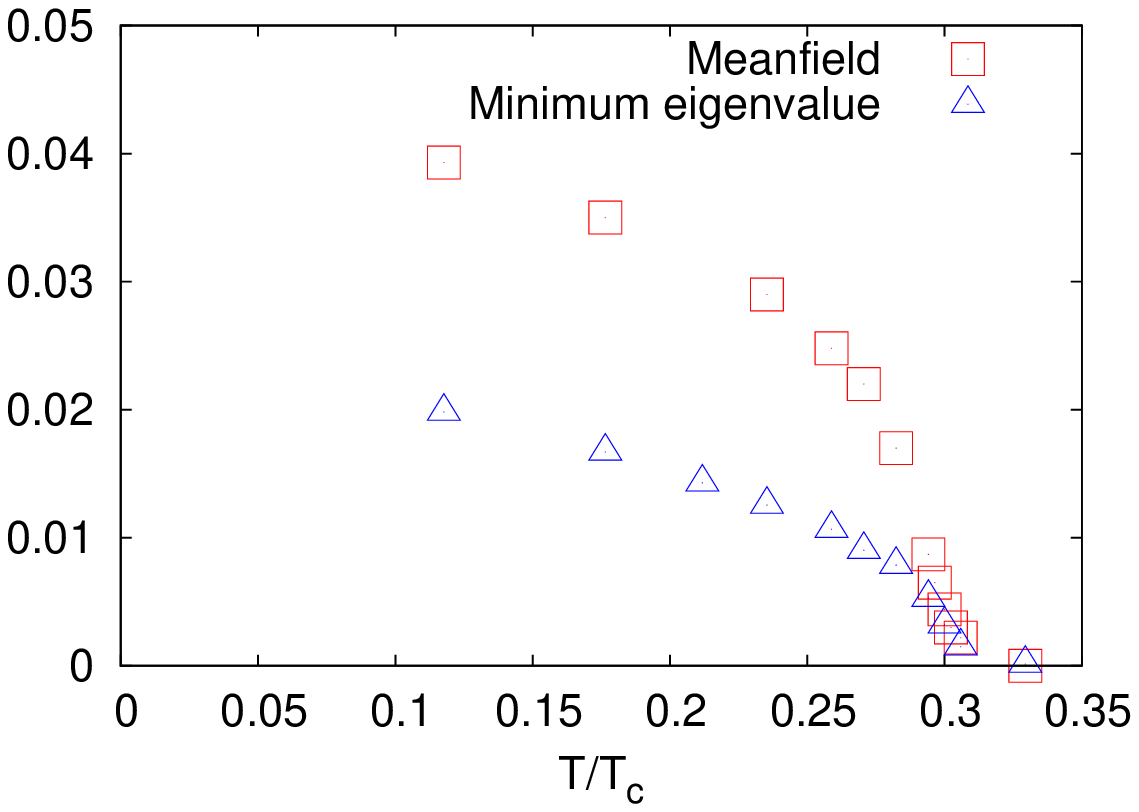}} 
    \end{tabular}
\end{center}
\caption{
(Color online) Temperature dependence of the extended $s$-wave order parameter 
at the midpoint of a side of the nanoisland and 
the minimum absolute eigenvalue. The critical temperature  is $T_{\rm c} = 0.085t$. 
\label{fig:supfig1}
 }
\end{figure}

\begin{figure}[t]
\begin{center}
     \begin{tabular}{p{ 0.7 \columnwidth}} 
      \resizebox{0.7 \columnwidth}{!}{\includegraphics{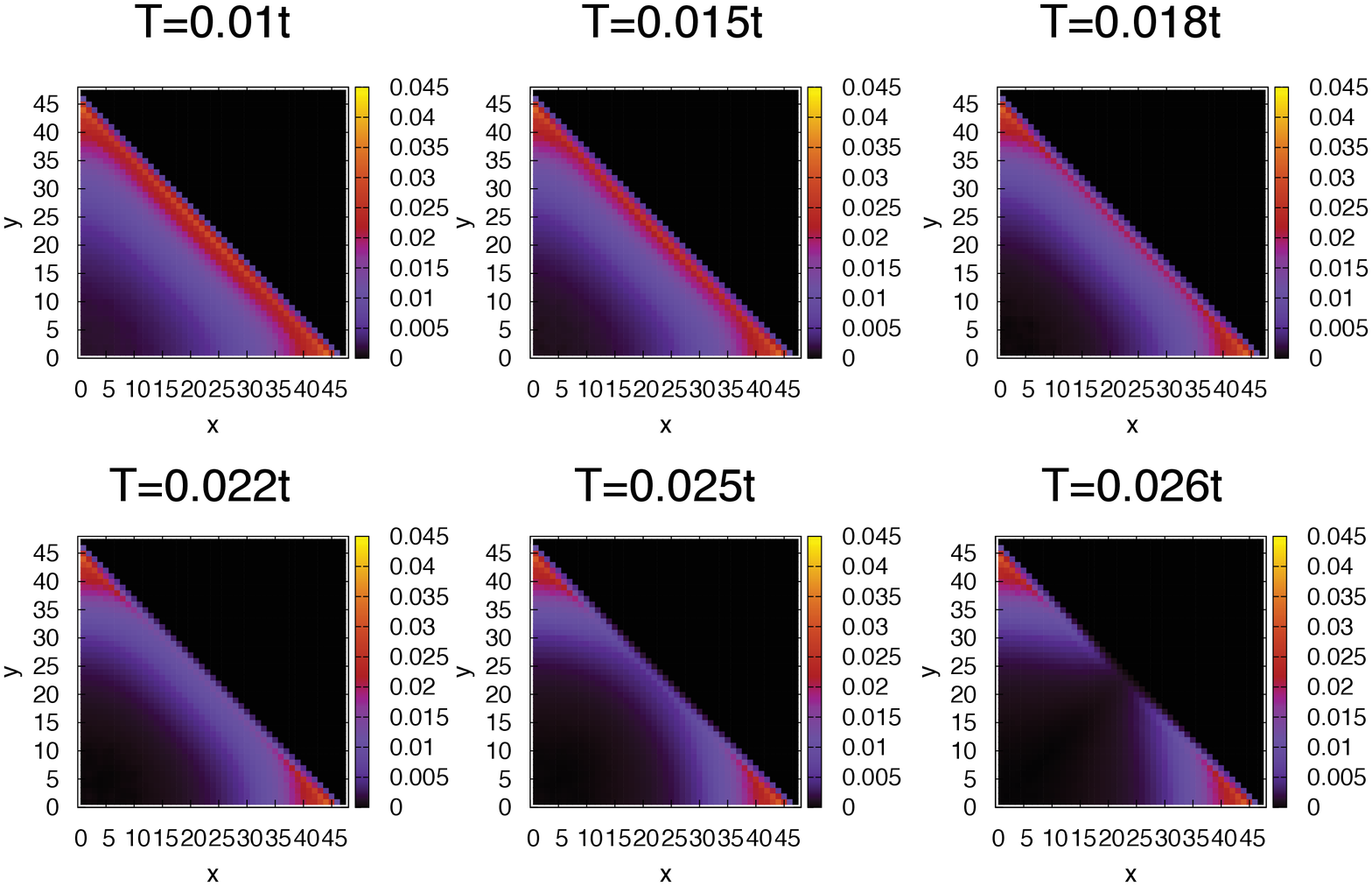}} 
    \end{tabular}
\end{center}
\caption{
(Color online) Temperature dependence of the extended $s$-wave order parameter. 
\label{fig:supfig2}
 }
\end{figure}

\begin{flushleft} 
{\bf S1. Temperature dependence of the extended $s$-wave order parameter}
\end{flushleft} 
Figure~\ref{fig:supfig1} shows the temperature dependence of both the extended $s$-wave 
order parameter at the midpoint of a side of the nanoisland and of
the minimum absolute eigenvalue. 
The zero-energy eigenvalues are split when the extended $s$-wave order parameter at the midpoint increases. 
As shown in Fig.~\ref{fig:supfig2}, one can clearly see the evolution of the extended $s$-wave order parameter along the side with decreasing temperature.

\begin{figure}[t]
\begin{center}
     \begin{tabular}{p{ 1 \columnwidth}} 
      \resizebox{1 \columnwidth}{!}{\includegraphics{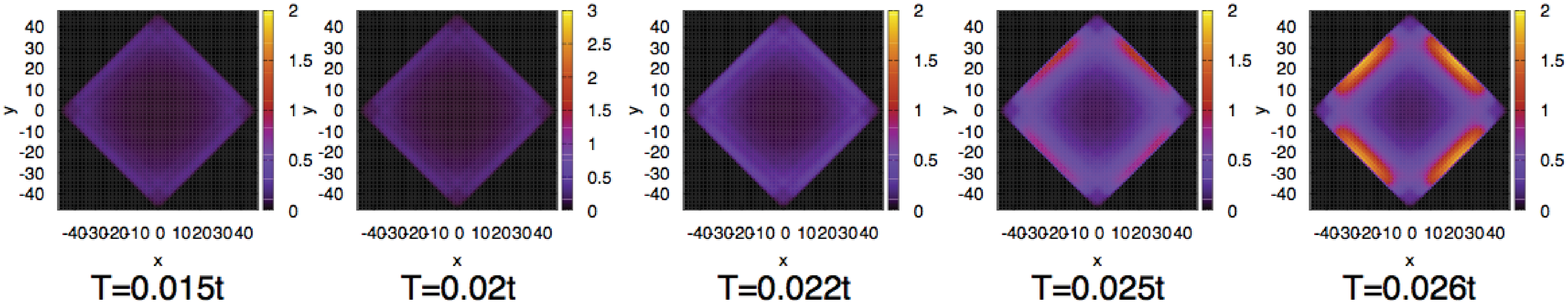}} 
    \end{tabular}
\end{center}
\caption{
(Color online) Temperature dependence of the zero-energy local density of states.
\label{fig:zeldos}
 }
\end{figure}

\begin{flushleft} 
{\bf S2. Temperature dependence of the zero-energy local density of states as a function of position on the nanoisland}
\end{flushleft} 
The local density of states at energy $E$ is calculated as 
\begin{align}
N(E,\Vec{r}_{i}) &= {1 \over \pi} \sum_{\gamma = 1}^{2 N} |u_{\gamma}(\Vec{r}_{i})|^{2} \frac{\eta}{(E-E_{\gamma})^{2} + \eta^{2}}, \label{eq:ldos}
\end{align}
where $\eta$ is a small parameter that serves to smear the density of states.
The eigenvalues and eigenvectors are calculated by the Sakurai-Sugiura (SS) method \cite{NagaiSSs,zParess}. 
We set $\eta = 0.01t$. 
Figure~\ref{fig:zeldos} shows the temperature dependence of the zero-energy local density of states as a function of position
on the nanoisland.

\begin{figure}[ht]
\begin{center}
     \begin{tabular}{p{ 0.5 \columnwidth}} 
      \resizebox{0.5 \columnwidth}{!}{\includegraphics{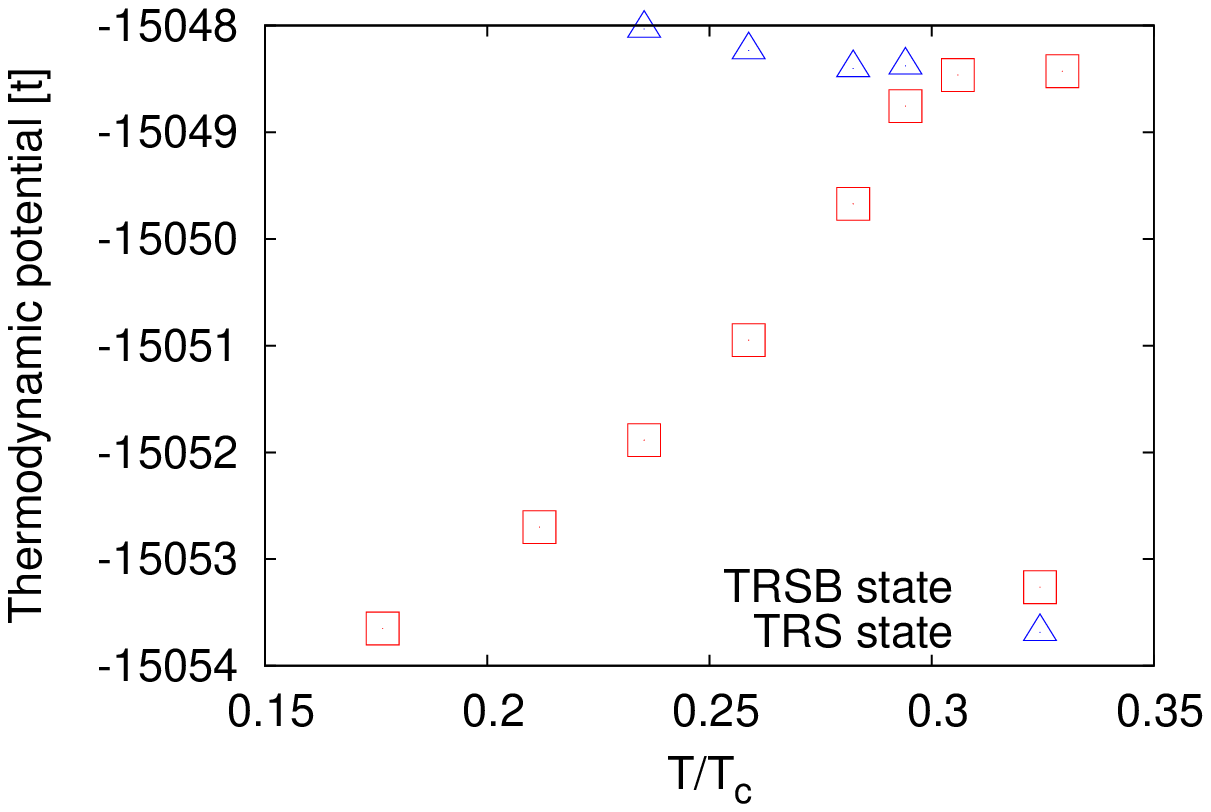}} 
    \end{tabular}
\end{center}
\caption{
(Color online) Temperature dependence of the thermodynamic potential for the two stable phases.
\label{fig:supfig3}
 }
\end{figure}

\begin{flushleft} 
{\bf S3. Temperature dependence of the thermodynamic potential}
\end{flushleft} 
The thermodynamic potential is given by Eq.~(4) in the main text.
If there are several stable phases, the phase with the lowest thermodynamic potential is realized.
In this system, there are two possible phases, the time-reversal symmetric phase and the time-reversal symmetry-broken phase. 
In our numerical calculation, the time-reversal symmetric phase can be obtained 
at all temperatures with the use of a real $d$-wave mean field as an initial guess. 
Figure~\ref{fig:supfig3} shows the temperature dependence of the thermodynamic potential, revealing that the thermodynamic potential in the time-reversal symmetry-broken phase is lower than that in the time-reversal symmetric phase.

\begin{flushleft} 
{\bf S4. Gap equations in real space}
\end{flushleft} 
We discuss the gap equations in real space.
We show that $\Delta_{d(s),i}$ is the $d$-wave (extended $s$-wave) order parameter. 
In real space, the gap equation can be written as 
\begin{align}
\Delta_{d(s),i}&= \sum_{j} \phi_{d(s)}(\Vec{r}_{i}-\Vec{r}_{j}) [\hat{\Delta} ]_{ij}, 
\end{align}
where 
\begin{align}
\phi_{d}(\Vec{r}_{i}-\Vec{r}_{j}) &\equiv (\delta(\Vec{r}_{i}-\Vec{r}_{j}-\hat{x}) + \delta(\Vec{r}_{i}-\Vec{r}_{j}+\hat{x}) 
-\delta(\Vec{r}_{i}-\Vec{r}_{j}-\hat{y}) - \delta(\Vec{r}_{i}-\Vec{r}_{j}+\hat{y}))/4, \\
\phi_{s}(\Vec{r}_{i}-\Vec{r}_{j}) &\equiv (\delta(\Vec{r}_{i}-\Vec{r}_{j}-\hat{x}) + \delta(\Vec{r}_{i}-\Vec{r}_{j}+\hat{x}) 
+\delta(\Vec{r}_{i}-\Vec{r}_{j}-\hat{y}) + \delta(\Vec{r}_{i}-\Vec{r}_{j}+\hat{y}))/4, \\
[\hat{\Delta} ]_{ij} &= V_{ij} \langle c_{i} c_{j} \rangle.
\end{align}
By introducing the Wigner representation expressed as 
\begin{align}
A(\Vec{r}_{i},\Vec{r}_{j}) &= \int d \Vec{k} e^{i \Vec{k} (\Vec{r}_{i} -\Vec{r}_{j})} A(\Vec{R},\Vec{k}),
\end{align}
where $\Vec{R} \equiv (\Vec{r}_{i}+\Vec{r}_{j})/2$,
the gap equation can be rewritten as 
\begin{align}
\Delta_{d(s),i}&= \sum_{j} \phi_{d(s)}(\Vec{r}_{i}-\Vec{r}_{j}) \int d \Vec{k} e^{i \Vec{k} (\Vec{r}_{i} -\Vec{r}_{j})} \Delta(\Vec{R},\Vec{k}),\\
&=  \int d \Vec{k}  \left( \sum_{j} e^{i \Vec{k} (\Vec{r}_{i} -\Vec{r}_{j})}  \phi_{d(s)}(\Vec{r}_{i}-\Vec{r}_{j}) \right)  \Delta(\Vec{R},\Vec{k}), \\
&= \int d \Vec{k}  \phi_{d(s)}(\Vec{k}) \Delta(\Vec{R},\Vec{k}).
\end{align}
The pairing function $\phi_{d(s)}(\Vec{k})$ is given as 
\begin{align}
\phi_{d}(\Vec{k}) &= (\cos k_{x} a - \cos k_{y} a)/2,\\
\phi_{s}(\Vec{k}) &= (\cos k_{x} a + \cos k_{y} a)/2,
\end{align}
with the lattice constant $a$. 
Thus, $\phi_{d}(\Vec{k})$ [$\phi_{s}(\Vec{k})$] is the $d$-wave [extended $s$-wave] order parameter, as 
$\phi_{d}(\Vec{k}) \sim -a^2(k_{x}^2 - k_{y}^2)/4$ \phantom{a} [$\phi_{s}(\Vec{k}) \sim 1 - a^2(k_{x}^2+k_{y}^2)/4$ ] in the long-wavelength limit.

\begin{figure}[t]
\begin{center}
     \begin{tabular}{p{ 1 \columnwidth}} 
      \resizebox{1 \columnwidth}{!}{\includegraphics{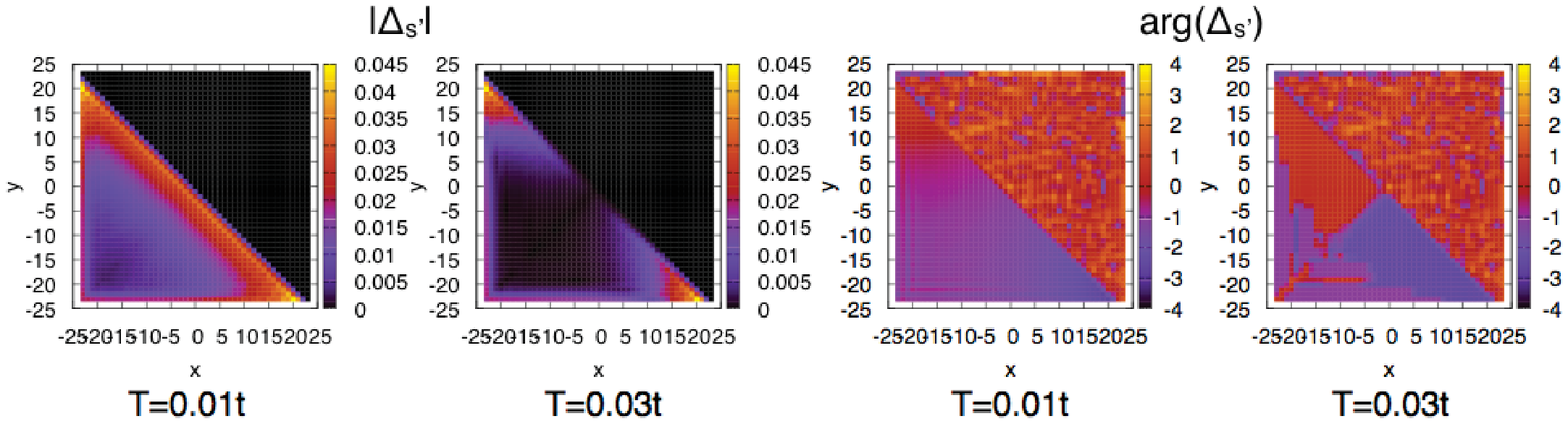}} 
    \end{tabular}
\end{center}
\caption{
(Color online) Temperature dependence of the extended $s$-wave order parameter 
in the triangular nanoisland.
\label{fig:supfig5}
 }
\end{figure}

\begin{figure}[t]
\begin{center}
     \begin{tabular}{p{ 0.4 \columnwidth}p{ 0.4 \columnwidth}} 
      (a)\resizebox{0.4 \columnwidth}{!}{\includegraphics{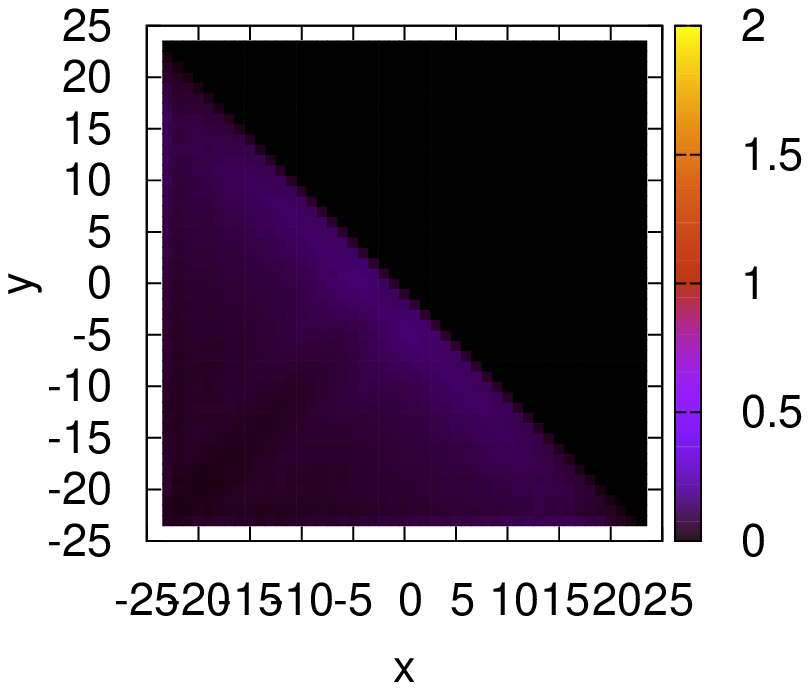}} 
      &
      (b)\resizebox{0.4 \columnwidth}{!}{\includegraphics{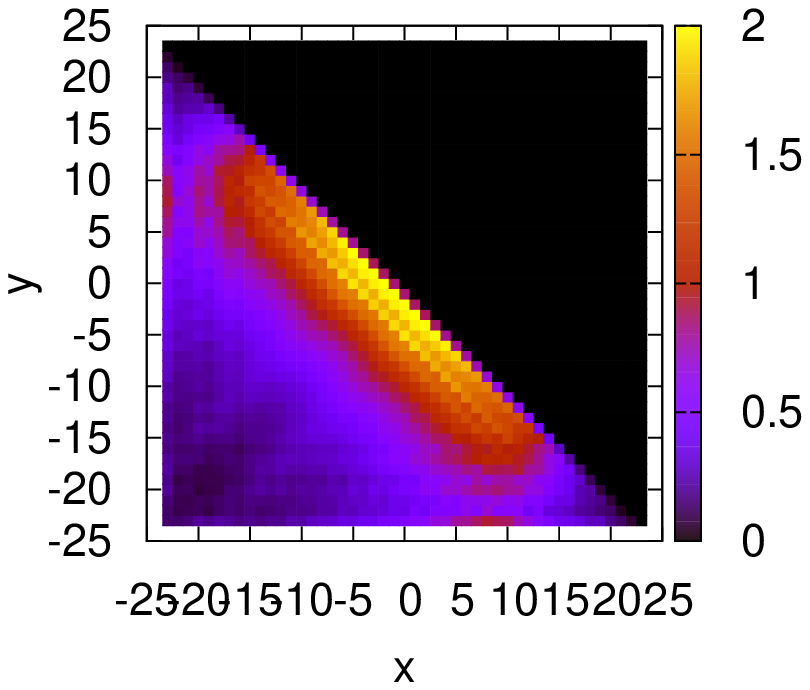}} 
    \end{tabular}
\end{center}
\caption{
(Color online) Zero-energy bound states in the triangular nanoisland for (a) $T = 0.01t$ and (b) $T=0.03t$.
\label{fig:supfig6}
 }
\end{figure}

\begin{flushleft} 
{\bf S5. Triangular nanoisland}
\end{flushleft} 
We show that the extended $s$-wave order parameter is induced only along a [110] surface. 
We consider a triangular nanoisland with the fabrication potential $V(x,y) = 500t$ in the region,
\begin{align}
y > -x + 2,
\end{align}
with the open boundary conditions on a $48 \times 48$ square lattice.
Here the origin is located at the center of the system. 
This shape includes both [100] and [110] surfaces.
Figure~\ref{fig:supfig5} presents the temperature dependence of the extended $s$-wave 
order parameter as a function of position in the triangular nanoisland. 
In Fig.~\ref{fig:supfig6} we show the zero-energy density of states calculated with Eq.~(\ref{eq:ldos}). 
The extended $s$-wave order parameter is induced only along the [110] surface, since the zero-energy bound states in a purely $d_{x^{2}-y^{2}}$-wave superconductor exist only on the [110] surface.

\begin{flushleft} 
{\bf S6. Topology and the zero-energy bound states}
\end{flushleft} 
The numerical calculations in the main text show that the presence 
of time-reversal symmetry is linked with the occurrence of 
zero-energy bound states on the [110] boundaries of a $d$-wave superconductor. 
To understand our results from the viewpoint of the bulk-edge 
correspondence, here we discuss a topological number in an analogous 
bulk system, according to the arguments in Refs.~\cite{SatoBs,Satoflats}. 

Let us consider the Bogoliubov-de Gennes (BdG) equations with single-orbital 
normal electrons and spin-singlet Cooper pairs in 2D momentum space. 
The BdG Hamiltonian is a $4 \times 4$ Hermitian matrix, 
$\mathcal{H}(\Vec{k})$. 
The wave vector $\Vec{k}$ is restricted to be in the first Brillouin zone. 
The time-reversal transformation is a $4 \times 4$ anti-unitary matrix, 
$\mathcal{T} = i \sigma^{y} \mathcal{K}$, where $\sigma^{y}$ is the $y$-component of the Pauli matrices in spin space,
and $\mathcal{K}$ is the complex 
conjugation operator, (the identity matrix in 
Nambu space is hidden). 
In the definition of $\mathcal{T}$, we take a basis representation 
in which the global constant phase of the pairing potential is zero. 
Time-reversal symmetry implies that 
$\mathcal{T}^{-1} \mathcal{H}(\Vec{k}) \mathcal{T} = \mathcal{H}(-\Vec{k})$. 
When the normal part of $\mathcal{H}(\Vec{k})$ is diagonal in spin 
space, $\mathcal{H}(\Vec{k})$ is of a block diagonal form, where 
each block is a $2 \times 2$ Hermitian matrix. 
Since one can find that the two blocks are not independent of each other, 
examining the eigenvalue problem in one block is sufficient for clarifying the 
topological character of our system~\cite{Satoflats}. 
Hereafter we focus on a special but closely related case to 
the calculations presented in the main text: 
the normal part is independent of spin. 
Thus, the normal electrons in the system do not intrinsically break 
time-reversal symmetry. 
The reduced $2 \times 2$ matrix reads
\begin{align}
\hat{H}(\Vec{k}) &= \left(\begin{array}{cc}
\epsilon(\Vec{k}) & \Delta(\Vec{k}) \\
\Delta(\Vec{k})^* & -\epsilon(\Vec{k})
\end{array}\right),
\end{align}
where $\epsilon(\Vec{k})$ is real, whereas $\Delta (\Vec{k})$ is complex. 
We also assume that $\epsilon(\Vec{k}) \neq |\Delta (\Vec{k})|$ 
for any $\Vec{k}$; 
this assumption is reasonable for typical superconducting systems. 
A straightforward calculation shows that $\mathcal{H}(\Vec{k})$ 
is invariant with respect to the time-reversal transformation if and 
only if $\Delta (\Vec{k})$ is real. 

Now, we focus on the reduced Hamiltonian. 
The ground state (i.e., the occupied band) of the system  
is well-defined with respect to any $\Vec{k}$ whenever the spectral gap, 
$2 \sqrt{\epsilon(\Vec{k})^{2} + |\Delta(\Vec{k})|^{2}}$, 
is not closed~\cite{Satoflats}. 
In a two-dimensional system, both $0=\epsilon(\Vec{k})$
and $0=|\Delta(\Vec{k})|$ describe one-dimensional geometrical objects. 
Therefore, their intersections must be points since 
$\epsilon (\Vec{k}) \neq |\Delta (\Vec{k})|$ for any $\Vec{k}$. 
The spectral gap of a $d_{x^{2}-y^{2}}$-wave superconductor, for example, 
is closed at four points on the Fermi surface 
since $\epsilon(\Vec{k})$ is zero on the Fermi surface and 
$\Delta(\Vec{k})$ is zero on the lines $|k_{x}| = |k_{y}|$. 
Thus, the spectral gap is closed 
only at finite numbers of points in momentum space.    
This fact indicates that we can find a closed one-dimensional path without 
passing through any of
the gap-closing points in the first Brillouin zone. 
Along such a path, a topological invariant in the ground state of the system
is well-defined. 
Among different choices of paths, we focus on an object suitable 
for revealing the presence of zero-energy bound states on the 
boundaries of the system, according to the concept of Ref.~\cite{Satoflats}:
we take a line along the direction perpendicular to momentum along the surface.
The $[100]$ surface of a $d_{xy}$-wave superconductor, for example, 
has the surface momentum $(0,k_{y})$ and then the corresponding path 
is parallel to $(k_{x},0)$. 

Let us consider the fundamental group \cite{Nakaharas}, $\pi_{1} (\mathcal{G})$, with 
the manifold $\mathcal{G}$ corresponding to the ground state of 
the system as we have a closed path in the first Brillouin zone. 
When a bulk system is invariant with respect to the time-reversal 
transformation, the reduced Hamiltonian is a $2\times 2$ real symmetric 
matrix. 
Therefore, the two-component ground-state eigenvector is parameterized 
by a single real number. 
It indicates that $\mathcal{G} = S^{1}$ when 
$\Delta^{\ast}(\Vec{k})=\Delta (\Vec{k})$. 
In contrast, if there is no time-reversal symmetry, the ground-state 
eigenvector is a nonzero two-component complex vector, 
$(z,\,w)$, with $z,\,w \in \mathbb{C}$. 
The nonzero-vector condition indicates $|z|^{2} + |w|^{2} \neq 0$, 
leading to $\mathbb{C}P^{1} \simeq S^{3} \simeq U(1)\times S^{2}$. 
Taking the ray representation into account, we need two real 
parameters to describe the ground-state eigenvector: it indicates 
that $\mathcal{G} = S^{2}$ when 
$\Delta^{\ast}(\Vec{k})\neq \Delta (\Vec{k})$. 
We find that 
$\pi_{1}(S^{1}) = \mathbb{Z}$ and $\pi_{1}(S^{2}) = 0$~\cite{Nakaharas}. 
Therefore, the presence of time-reversal symmetry ensures a nontrivial 
fundamental group. 
In other words, zero-energy bound states on a one-dimensional surface of 
a two-dimensional superconductor are protected by time-reversal symmetry. 
When time-reversal symmetry in the superconducting order is broken, 
the fundamental group associated with the one-dimensional surface is trivial. 
Thus, the absence of zero-energy bound states shown in the main text 
can be understood from the topological point of view. 

A more detailed feature of the zero-energy bound states (e.g., dispersion 
relation) can be found by a one-dimensional topological number~\cite{Satoflats}. 
We briefly review this argument below. 
In a uniform system with time-reversal symmetry, the reduced 
Hamiltonian is a $2 \times 2$ real symmetric matrix. 
Repeating the above arguments, the ground-state eigenvector can be expressed as
\begin{align}
\Vec{\psi}(\Vec{k}) &= \left(\begin{array}{c}
\cos \alpha(\Vec{k}) \\
\sin \alpha(\Vec{k})
\end{array}\right)
\end{align}
with a single variable $\alpha(\Vec{k})$. 
The variable $\alpha(\Vec{k})$ is defined in the range 
$0 \le \alpha(\Vec{k}) \le \pi$, where 
$-\Vec{\psi}(\Vec{k})$ is identical to $\Vec{\psi}(\Vec{k})$ for $\pi \le \alpha(\Vec{k}) \le 2\pi$.
Thus, the eigenvector $\Vec{\psi}(\Vec{k})$ is equivalent to a circle 
defined by $(\cos 2 \alpha(\Vec{k}), \sin 2 \alpha(\Vec{k}))$. 
We consider the momentum $\Vec{k} = (k_{x},k_{y})$. 
By fixing the $y$ component $k_{y}=k_{y}^{0}$, the range of the 
variable $k_{x}$ is $- \pi \le k_{x} \le \pi$. 
If the path $\Vec{k} = (k_{x},k_{y}^{0})$ in momentum space 
does not cross the nodes, the winding number associated with $\pi_{1}(S^{1})$
is regarded as a one-dimensional topological invariant. 
According to Ref.~\cite{Satoflats}, the winding number is 
\begin{align}
w(k_{y}^{0}) &= \frac{1}{2}
{\rm sgn} \: \left[ 
\frac{\partial \epsilon(-k_{x}^{0},k_{y}^{0})}{\partial k_{x}} \right] 
\left(
{\rm sgn} \: \left[ \Delta(-k_{x}^{0},k_{y}^{0}) \right]-{\rm sgn} \: \left[ \Delta(k_{x}^{0},k_{y}^{0}) \right]
\right),
\end{align}
where $(\pm k_{x}^{0},k_{y}^{0})$ denote the intersection points between the integration path with the fixed $k_{y}^{0}$ and the Fermi surface. 
Thus, if the gap function $\Delta(\Vec{k})$ satisfies 
\begin{align}
\Delta(-k_{x}^{0},k_{y}^{0}) \Delta(k_{x}^{0},k_{y}^{0}) < 0, \label{eq:cond}
\end{align}
then the one-dimensional topological invariant is finite and the zero-energy bound states 
exist because of the bulk-edge correspondence. 
Varying $k_{y}^{0}$ in the range with no gap closing, we can obtain the surface 
dispersion relation of the zero-energy bound states~\cite{Satoflats}. 

In a pure $d_{x^{2}-y^{2}}$ superconductor, the gap function is expressed as 
\begin{align}
\Delta_{d}(\Vec{k}) &= \frac{\Delta_{0}}{2} (\cos (k_{x}) - \cos(k_{y})).
\end{align}
The zero-energy bound states exist on the [110] surface, since the nodes are located along the lines $|k_{x}| = |k_{y}|$.  
In the $d_{x^{2}-y^{2}}$ superconductor with uniform extended $s$-wave pairs, 
the gap function can be expressed as 
\begin{align}
\Delta_{d+s'}(\Vec{k}) &= \frac{\Delta_{0}}{2} ( \cos (k_{x}) - (1 - 2r) \cos(k_{y})).
\end{align}
Here $r$ denotes the ratio between the amplitudes of the $d$- and $s'$-wave gap functions. 
The nodes are located on the lines which satisfy the relation,
\begin{align}
\cos (k_{y}) &= \frac{1}{1-2r} \cos(k_{x}).
\end{align}
If the Fermi surface is not too small and the $d_{x^{2}-y^{2}}$ component is larger than the extended $s$-wave one $(r < 0.5)$, there are always zero-energy bound states on the [110] surface.

\begin{figure}[t]
\begin{center}
     \begin{tabular}{p{ 0.24 \columnwidth}p{ 0.24 \columnwidth}p{ 0.24 \columnwidth}p{ 0.24 \columnwidth}} 
      (a)\resizebox{0.24 \columnwidth}{!}{\includegraphics{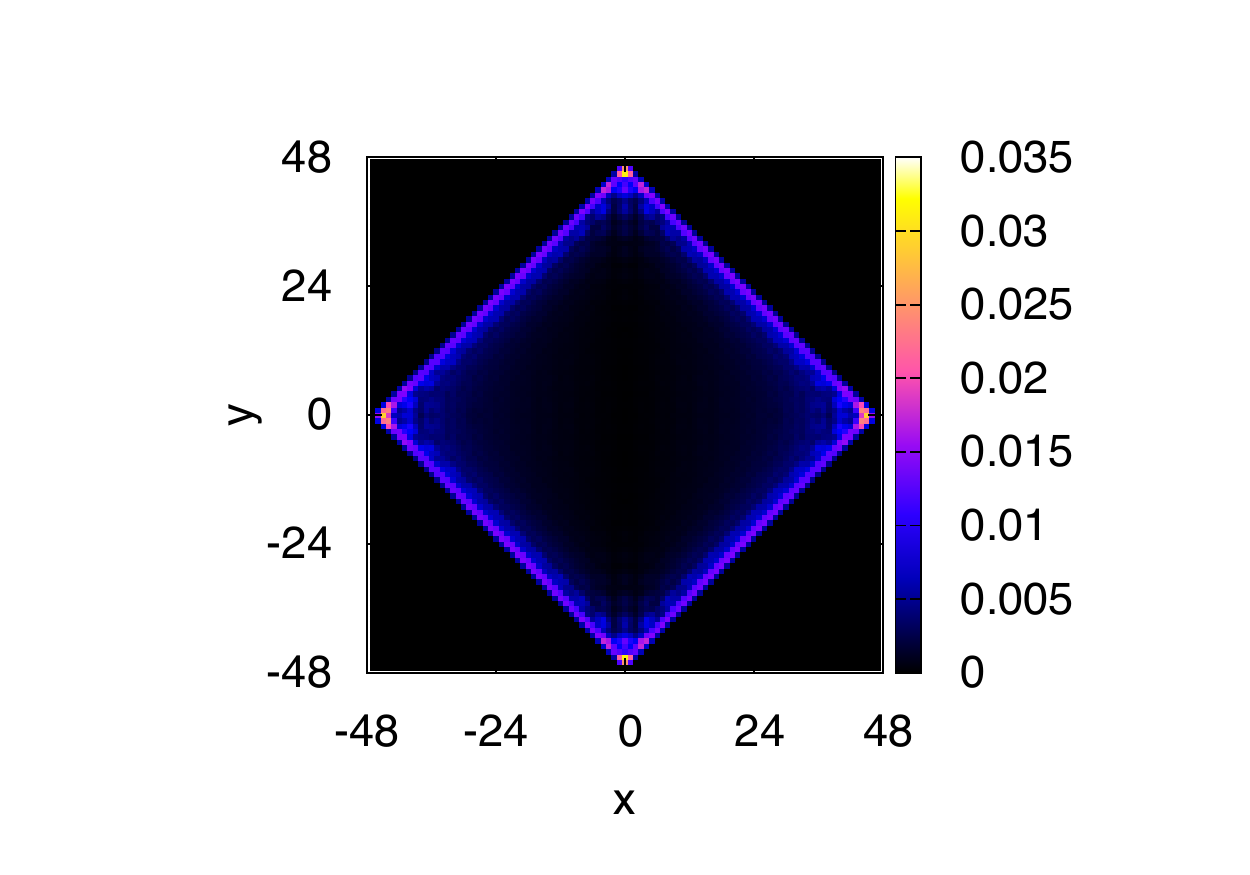}} 
      &
      (b)\resizebox{0.24 \columnwidth}{!}{\includegraphics{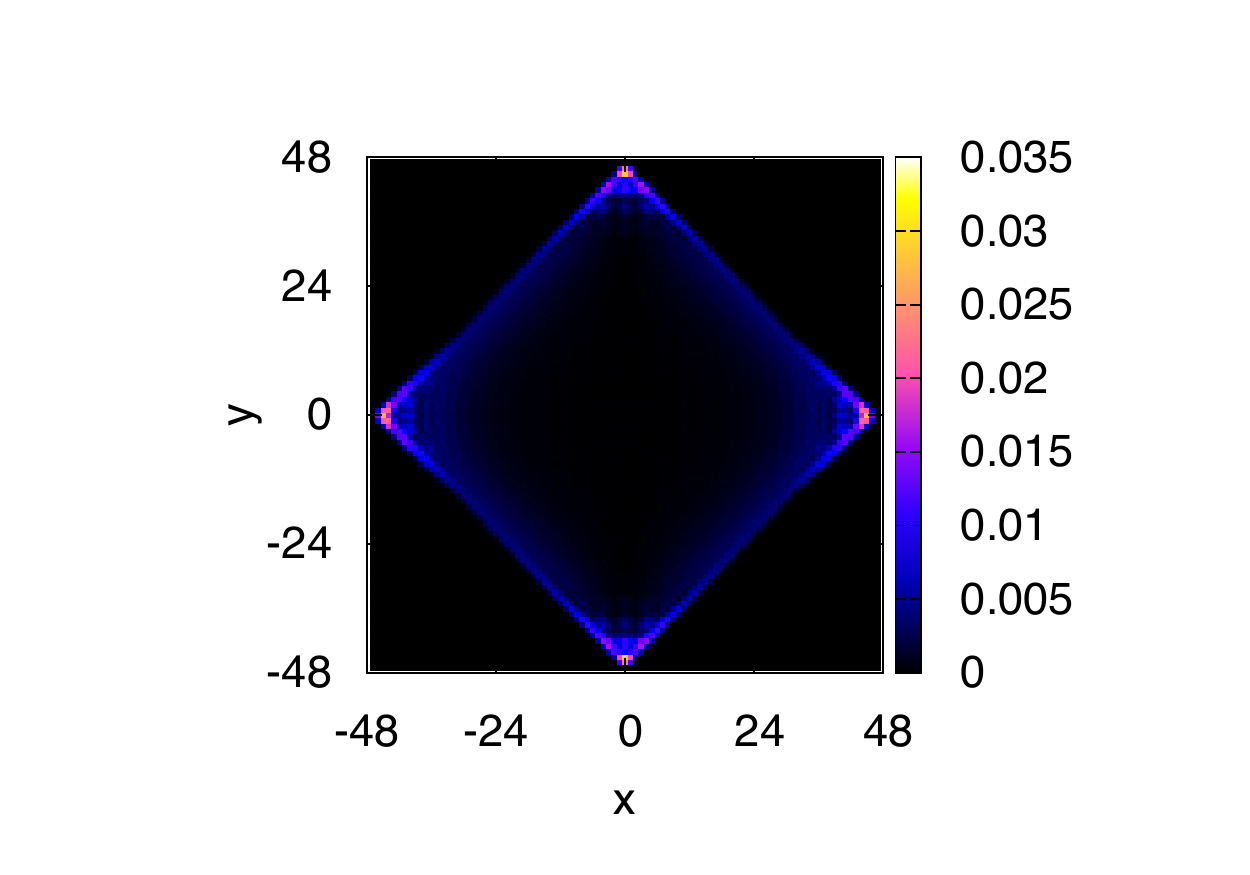}} 
            &
      (c)\resizebox{0.24 \columnwidth}{!}{\includegraphics{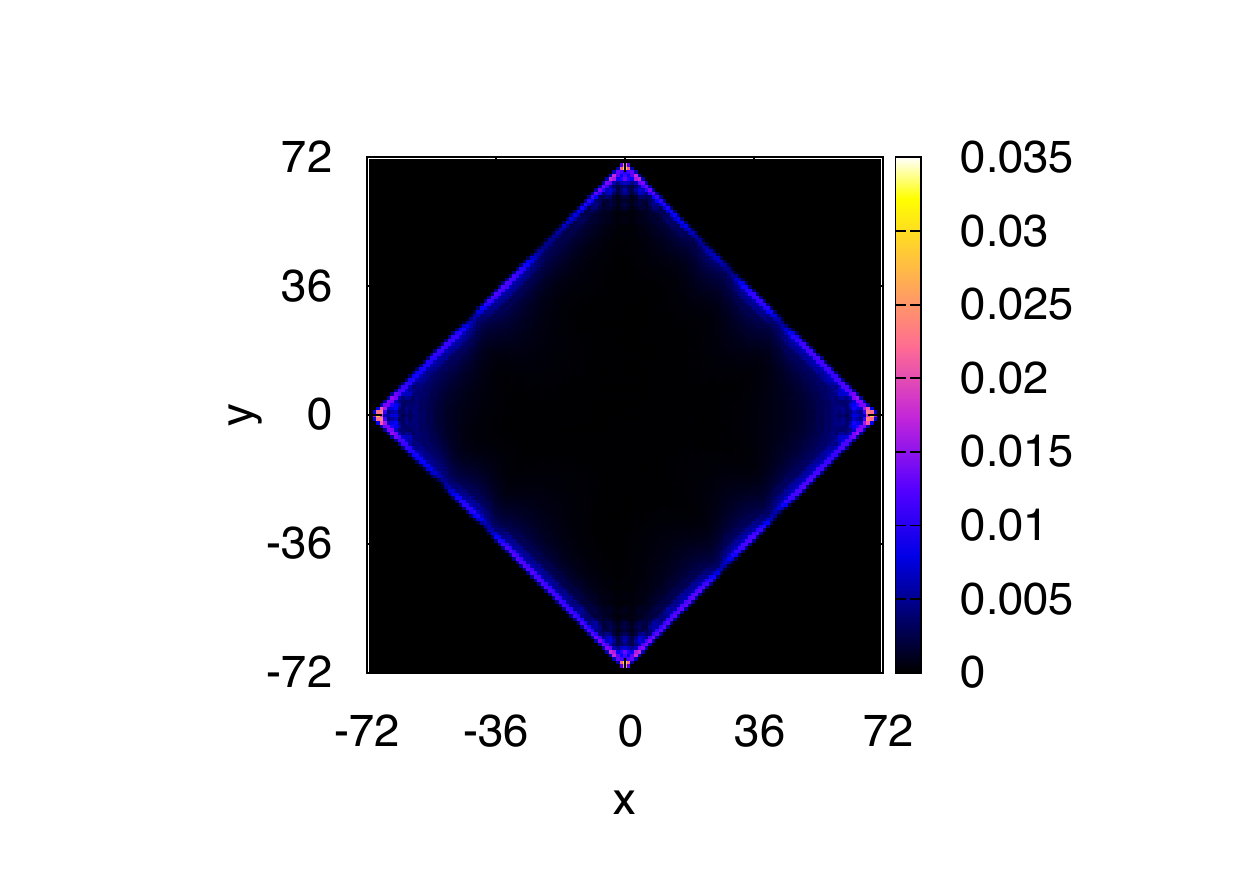}} 
            &
      (d)\resizebox{0.24 \columnwidth}{!}{\includegraphics{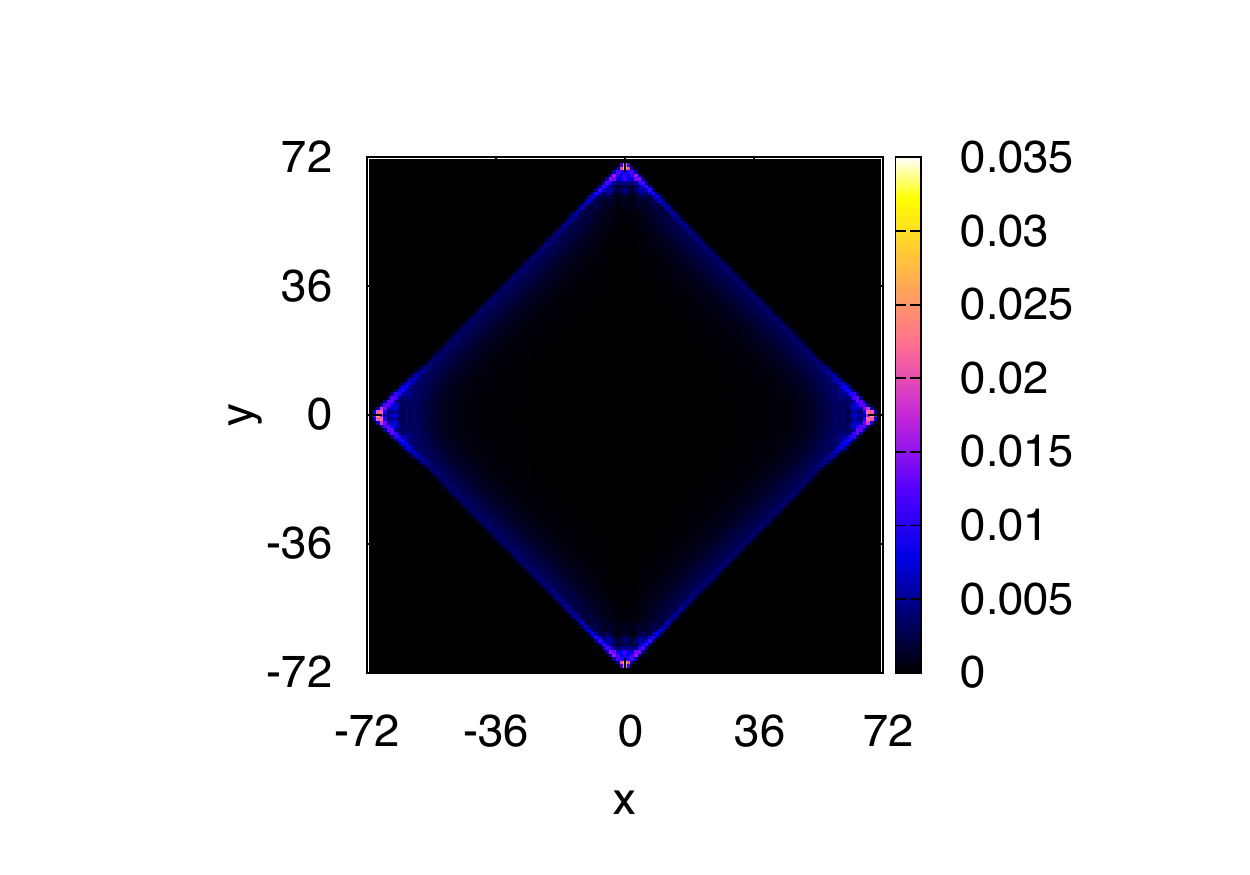}} 
    \end{tabular}
\end{center}
\caption{
(Color online) Amplitude of the $p_{x}$-wave order parameter. In (a) and (b), the system size is $96 \times 96$ lattice sites, while it is  $144 \times 144$ sites in (c) and (d).
The temperature is $T = 0.02t$ for (a) and (c) and $T = 0.03t$ for (b) and (d).
\label{fig:supfig7}
 }
\end{figure}

\begin{flushleft} 
{\bf S7. $p$-wave order parameters}
\end{flushleft} 
Finally, we have studied whether the order parameters with $p_{x}$- and $p_{y}$-wave symmetry as defined by
\begin{align}
\Delta_{p_{x},i} &= (\Delta_{\hat{x},i} - \Delta_{-\hat{x},i})/2,\\
\Delta_{p_{y},i} &= (\Delta_{\hat{y},i} - \Delta_{-\hat{y},i})/2,
\end{align}
are induced in nanoislands. In Fig.~\ref{fig:supfig7} we show the spatial distribution of the $p_{x}$-wave order parameter amplitude $|\Delta_{p_{x},i}|$ for the nanoisland with $96 \times 96$ lattice sites [(a) and (b)] and $144 \times 144$ sites [(c) and (d)], for $\mu=1.5t$ and $U=-2t$ at temperature $T = 0.02t<T_{\rm cs}$ [(a) and (c)] and $T = 0.03t>T_{\rm cs}$ [(b) and (d)]. We find that the $p_{y}$-wave amplitude, $|\Delta_{p_{y},i}|$, is exactly the same as the $p_{x}$-wave amplitude. It can be seen in Fig.~\ref{fig:supfig7} that $|\Delta_{p_{x},i}|$ is induced along the surfaces within a range narrower than the regions where the extended $s'$-wave order is induced, although right on the surfaces it can be comparable to $|\Delta_{s',i}|$. Most significantly, however, no fundamental change occurs in $|\Delta_{p_{x},i}|$ and $|\Delta_{p_{y},i}|$ as the temperature is lowered from above ($T = 0.03t$) to below ($T = 0.02t$) the TRS-breaking transition temperature. Therefore, we conclude that the $p_{x}$- and $p_{y}$-wave order parameters induced close to the surfaces are not related to the spontaneous breaking of TRS.

\end{document}